\documentstyle[prb,tighten,aps,epsf]{revtex}
\begin{document}
%\draft
\title{Current and power spectrum in a magnetic tunnel device with  an atomic size spacer}
\author{Bogdan R. Bu{\l}ka}
%\email{bulka@ifmpan.poznan.pl}
%\affiliation
\address
{Institute of Molecular Physics, Polish Academy of
Sciences, ul.Smoluchowskiego 17, 60-179 Pozna\'n, Poland}

\date{Received \hspace{5mm} }
\date{\today}

\maketitle\draft
\begin{abstract}
Current and its noise in a ferromagnetic double tunnel barrier
device with
 a small spacer particle were studied in the framework of the sequential
 tunneling approach.
 Analytical formulae were derived for electron tunneling through the spacer
 particle containing
 only a single energy level.
It was shown that Coulomb interactions of electrons with a
different spin orientation  lead to an increase of the tunnel
magnetoresistance. Interactions can also be responsible for the
negative differential resistance. A current noise study showed,
which relaxation processes can enhance or reduce fluctuations
leading either to a super-Poissonian or a sub-Poissonian shot
noise.
\end{abstract}
% \maketitle

\section{Introduction}

Recent interest in single-electron tunneling in ferromagnetic
double tunnel junctions is stimulated by expected potential
applications at microelectronics and by new phenomena observed in
such systems.~\cite{fsetex,fsette} In order to have a device
operating at room temperature the single electron charging energy
$E_c=e^2/2C$ should be much larger than the thermal energy $k_BT$.
It can be achieved decreasing the capacitance $C$ of the metallic
spacer, which is proportional to its size. In a small metallic
spacer a discreteness of the energy spectrum can be relevant and a
separation of energy levels $\Delta E \approx k_BT$. Such the
situation was studied numerically just recently.~\cite{prb}

In the present paper we would like to investigate sequential
tunneling in an extreme case, when the spacer particle has only a
single electron level available for the tunneling process. This
simplified model gives us possibility to gain a better insight
into spin dependent tunneling processes and to solve the problem
analytically. We will show that Coulomb interactions between
electrons with different spins can lead to new effects. In some
circumstances due to the Coulomb blockade effect the device can
operate as a diode, in others it can show the negative
differential resistance (NDR). The power spectrum analysis will be
performed to understand correlations between currents for
electrons of different spins and the transition from the
sub-Poissonian to the super-Poissonian current noise in the
ferromagnetic device.

\section{Model and general derivations}

Let us specify the system considered in detail. The separation
between the ferromagnetic metallic electrodes is large and
therefore, there is no direct electron tunneling between them. The
electronic transport can be only via electronic states of the
spacer particle placed between the electrodes. The particle can be
a molecule (e.g. $C_{60}$), or a semiconductor quantum dot, in
which the relevant energies are $\Delta E, E_c \gg k_BT$. For a
small applied voltage $V$ ($eV\ll \Delta E, E_c$)  electronic
transport is only through a single electronic level $E_0$. Such
the model was considered for a nonmagnetic device in
Ref.[\onlinecite{gla,naz,her}] and we generalize it for a
ferromagnetic case including tunneling channels for electrons with
opposite spin directions. The tunneling process for an electron
with spin $\sigma$ through the left ($j=1$) and the right ($j=2$)
junction is described by the net tunneling rates
$\gamma_{j\sigma}$, which are assumed to be small $\hbar
\gamma_{j\sigma}\ll k_BT$. This relation implies that the
corresponding tunnel resistances $R_{j\sigma}$ are much larger
than the quantum resistance $R_Q = h/2e^2$ and electronic
transport can be described within the sequential tunneling
approach.~\cite{schon,gla,naz,her} Since $\Delta E$ is large, the
tunneling process can be considered elastic (there is no
thermalization of electrons on the spacer particle, which was
usually assumed in the single electron transistor with a large
metallic grain)~\cite{schon,prb}. We also neglect fluctuations of
the position of the electronic level $E_0$, which can be caused by
thermal and electrostatic fluctuations of the environment.

Our model seems to be familiar to that considered recently for the
Kondo effect in quantum dots.~\cite{kondo} A condition for a
development of the Kondo resonance is a buildup of many-body
correlations between the dot and the electrodes, which can be
achieved when electronic waves are coherently scattered on a
magnetic impurity. It is in contrast to the present situation,
where coupling between the particle and the electrodes is weak and
electron tunneling events are uncorrelated and incoherent.

\subsection{Stationary currents}

Electronic transport is governed by the master equation
\begin{eqnarray}\label{1}
\frac{d}{dt} \left[\begin{array}{c}
p_{\uparrow}\\p_{\downarrow}\\p_0\end{array}\right]=\hat{M}\left[\begin{array}{c}
p_{\uparrow}\\p_{\downarrow}\\p_0\end{array}\right]\;,
\end{eqnarray}
where $p_{\uparrow}$ and $p_{\downarrow}$ denotes the probability
to find an electron with the spin $\sigma=\uparrow$ and
$\downarrow$, $p_0$ - the probability for an empty state $E_0$. Of
course, the total probability $p_{\uparrow}+p_{\downarrow}+p_0=1$.
The matrix $\hat{M}$ is given by
\begin{eqnarray}\label{2}
\hat{M}=\left[\begin{array}{ccc}
-\Gamma^-_{\uparrow}&0&\Gamma^+_{\uparrow}\\
0&-\Gamma^-_{\downarrow}&\Gamma^+_{\downarrow}\\
\Gamma^-_{\uparrow}&\Gamma^-_{\downarrow}&-\Gamma^+_{\uparrow}-\Gamma^+_{\downarrow}
\end{array}\right]\;,
\end{eqnarray}
where
$\Gamma^{\pm}_{\sigma}=\Gamma^{\pm}_{1\sigma}+\Gamma^{\pm}_{2\sigma}$,
$\Gamma^{\pm}_{j\sigma}= f^{\pm}_j\gamma_{j\sigma}$ are the total
tunneling rates to ($+$) and off ($-$) the particle level $E_0$,
$f^{\pm}_j=\{1+\exp[\pm(E_0-E_F-(-1)^jeV_j)/k_BT]\}^{-1}$. The
voltage $V$ is applied to the left electrode and the voltage drop
across the left and the right junction is $V_1=C_2V/C$ and
$V_2=C_1V/C$, respectively. Here, $C_j$ denotes the capacitance of
the $j$-th tunnel junction and $C=C_1+C_2$.

At the stationary state the probability $p_{\sigma}$ and $p_0$ are
determined from the master equation (\ref{1}) with the left hand
side equal to zero, and the result is
\begin{equation}\label{3}
p_{\sigma}=\frac{\Gamma^+_{\sigma}\Gamma^-_{-\sigma}}{\gamma_{\uparrow}\gamma_{\downarrow}
-\Gamma^+_{\uparrow}\Gamma^+_{\downarrow}}\;,\;\;\;
p_0=\frac{\Gamma^-_{\uparrow}\Gamma^-_{\downarrow}}{\gamma_{\uparrow}\gamma_{\downarrow}
-\Gamma^+_{\uparrow}\Gamma^+_{\downarrow}}\;,
\end{equation}
where $\gamma_{\sigma}=\gamma_{1\sigma}+\gamma_{2\sigma}$. The
current through the left junction for electrons with the spin
$\sigma$ is the difference of the tunneling current flowing to (+)
and from ($-$) the particle
\begin{eqnarray}\label{4}
I_{1\sigma}\equiv I^+_{1\sigma}-I^-_{1\sigma} =
-e\left[\Gamma^+_{1\sigma}p_0-\Gamma^-_{1\sigma}p_{\sigma}\right]\nonumber\\
= -e(f^+_1-f^+_2)
\frac{\gamma^+_{1\sigma}\gamma^+_{2\sigma}\Gamma^-_{-\sigma}}{\gamma_{\uparrow}\gamma_{\downarrow}
-\Gamma^+_{\uparrow}\Gamma^+_{\downarrow}}\;.
\end{eqnarray}
Since there are no electronic relaxation processes on the
particle, it results from the current conservation rule that
$I_{1\sigma}=I_{2\sigma}$ for each electronic channel.

In magnetic tunnel junctions the resistance depends on the
relative configuration of magnetic moments in the electrodes and
this effect is known as the tunnel magnetoresistance (TMR). The
value of TMR is given by the ratio $TMR=(I_P-I_{AP})/I_{AP}$,
where $I_P$ and $I_{AP}$ are the tunneling currents in the
parallel (P) and the antiparallel (AP) configuration of the
magnetic moments in the electrodes.  It is convenient to express
the tunneling rate coefficients  in the form
$\gamma_{1\sigma}=\gamma_0(1\pm P_1)$ and
$\gamma_{2\sigma}=\gamma_0\alpha(1\pm P_2)$, where the sign $+$
($-$) corresponds to the spin $\sigma=\uparrow$ ($\downarrow$),
$P_1$ and $P_2$ is the magnetic polarization of the left and the
right electrode, $\alpha$ denotes the asymmetry between the
potential barriers. Using Eq.(\ref{4}) one gets
%\begin{widetext}
\begin{equation}\label{5}
TMR=
 \frac{(1-f^+_1f^+_2)4\alpha P_1P_2}
{(1+\alpha)^2-(P_1+\alpha P_2)^2-(f^+_1+\alpha f^+_2)^2
+(P_1f^+_1+\alpha f^+_2P_2)^2}\;.
\end{equation}
%\end{widetext}

For comparison we present the results for noninteracting
electrons,
 i.e when the single electron charging energy $E_c=0$. In this limit the double
 occupancy of the level $E_0$ is allowed.  The current through the left junctions
 for electrons with the spin $\sigma$ is then
\begin{equation}\label{5b}
I^0_{1\sigma}=
-e(f^+_1-f^+_2)\frac{\gamma_{1\sigma}\gamma_{2\sigma}}{\gamma_{\sigma}}
\end{equation}
and TMR
\begin{equation}\label{5c}
TMR^0= \frac{4\alpha P_1P_2} {(1+\alpha)^2-(P_1+\alpha P_2)^2}\;.
\end{equation}
Comparison of both the expressions for TMR [Eq.(\ref{5}) and
(\ref{5c})] shows that Coulomb interactions can significantly
increase the value of the magnetoresistance.

\subsection{Fluctuations}

Fluctuations in the system are studied within the
generation-recombination approach for multi-electron
channels.~\cite{vliet,kor,prb} The Fourier transform of the
correlation function of the quantity $X$ can be expressed
as~\cite{vliet,kor}
\begin{eqnarray}\label{7}
S_{XX}(\omega)\equiv 2\int_{-\infty}^{\infty}dt e^{i\omega
t}\left[\langle X(t)X(0)\rangle-\langle X
\rangle^2\right]\nonumber\\
=4\sum_{n,m}X_{n}\left[P(n,m;\omega)-\frac{p_n}{i\omega}\right]X_mp_m,\;\;\;
\end{eqnarray}
where $p_m$ is the stationary value of the probability $\hat{p}$
at the state $m$ [given by Eq.(\ref{3})], $X_m$ is the value of
$X$ at this state. The conditional probability $P(n,m;t)$ to find
the system in the state $n$ at time $t$, if it was in the initial
state $m$ at $t=0$, satisfies the master equation
(\ref{1}),~\cite{vliet,kor} and its Fourier transform is given by
$P(n,m;\omega)=[i\omega-\hat{M}]^{-1}_{nm}$. The elements of the
Green's function $G(n,m;\omega)\equiv
[i\omega-\hat{M}]^{-1}_{nm}-p_n/i\omega$ can be determined
directly by  matrix inversion and the result is
\begin{equation}\label{8}
\hat{G}(\omega)=\frac{\hat{A}^+}{i\omega-\lambda_+}-\frac{\hat{A}^-}{i\omega-\lambda_-}\;,
\end{equation}
where $\lambda_{\pm}=(-\gamma_{\uparrow}-\gamma_{\downarrow}\pm
\Delta)/2$ are the nonzero eigenvalues of the matrix $\hat{M}$,
$\Delta=\sqrt{(\gamma_{\uparrow}-\gamma_{\downarrow})^2+4\Gamma^+_{\uparrow}
\Gamma^+_{\downarrow}}$,
\begin{eqnarray}\label{9}
\hat{A}^r=\frac{1}{D} \left[\begin{array}{ccc}
\Gamma^-_{\uparrow}a^r_{\uparrow,\uparrow}&\Gamma^+_{\uparrow}a^r_{\uparrow,\downarrow}
&\Gamma^+_{\uparrow}a^r_{\uparrow,0}\\
\Gamma^-_{\downarrow}a^r_{\downarrow,\uparrow}&
\Gamma^+_{\downarrow}a^r_{\downarrow,\downarrow}&
\Gamma^+_{\downarrow}a^r_{\downarrow,0}\\
\Gamma^-_{\uparrow}a^r_{\uparrow,0}&
\Gamma^-_{\downarrow}a^r_{\downarrow,0}&
-\Gamma^+_{\uparrow}a^r_{\uparrow,0}-\Gamma^+_{\downarrow}a^r_{\downarrow,0}
\end{array}\right]
\end{eqnarray}
corresponding to $\lambda_r$ ($r=\pm$),
$a^r_{\sigma,\sigma}=\lambda_r\gamma_{-\sigma}+\gamma_{-\sigma}^2+\Gamma^+_{\uparrow}
\Gamma^+_{\downarrow}$,
$a^r_{\sigma,-\sigma}=-\Gamma^-_{-\sigma}(\lambda_r+\gamma_{\uparrow}+
\gamma_{\downarrow})$,
$a^r_{\sigma,0}=-(\lambda_r+\gamma_{-\sigma})\Gamma^-_{-\sigma}+
\Gamma^+_{-\sigma}\Gamma^-_{\sigma}$, and $D=\Delta
(\gamma_{\uparrow}\gamma_{\downarrow}-\Gamma^+_{\uparrow}\Gamma^+_{\downarrow})$.
The Green's function (\ref{8}) is not singular for $\omega\to 0$
and therefore, one can easily separate the amplitudes of the noise
resulting from fluctuation processes characterized by the
relaxation time $\tau_r=-1/\lambda_r$.

The fluctuations of the charge and the spin are expressed as
\begin{eqnarray}\label{10}
S_{NN}(\omega)=4e^2\sum_{\sigma,\sigma'}G_{\sigma\sigma'}(\omega)p_{\sigma'}=
\nonumber\\
\frac{4e^2\Gamma^-_{\uparrow}\Gamma^-_{\downarrow}}{\Delta(\gamma_{\uparrow}\gamma_{\downarrow}
-\Gamma^+_{\uparrow}\Gamma^+_{\downarrow})^2}\sum_{\sigma,r}r
\frac{\Gamma^+_{-\sigma}\Gamma^-_{\sigma}(\lambda_r
+\Gamma^-_{\sigma})}{i\omega-\lambda_r}\;,
\end{eqnarray}
\begin{eqnarray}\label{11}
S_{MM}(\omega)=4\frac{\mu_B^2}{4}\sum_{\sigma,\sigma'}\sigma\sigma'G_{\sigma\sigma'}(\omega)
p_{\sigma'}= \nonumber\\
\frac{\mu_B^2\Gamma^-_{\uparrow}\Gamma^-_{\downarrow}}{\Delta(\gamma_{\uparrow}\gamma_{\downarrow}
-\Gamma^+_{\uparrow}\Gamma^+_{\downarrow})^2}\sum_{\sigma,r}r
\frac{\Gamma^+_{-\sigma}(\gamma_{\sigma}+\Gamma^+_{\sigma})(\lambda_r
+\gamma_{\sigma}+\Gamma^+_{\sigma})}{i\omega-\lambda_r}\;,
\end{eqnarray}
where $\mu_B$ is the Bohr magneton.

The correlations between the currents $I_{j\sigma}$ and
$I_{j'\sigma'}$ in the tunnel junction $j$ and $j'$ for the
electrons with the spin $\sigma$ and $\sigma'$ are described by
the power spectrum~\cite{kor}
\begin{equation}\label{12}
S_{I_{j\sigma}I_{j'\sigma'}}(\omega)=\delta_{jj'}\delta_{\sigma\sigma'}
S^{Sch}_{j\sigma}+S^c_{I_{j\sigma}I_{j'\sigma'}}(\omega)\;,
\end{equation}
where
\begin{equation}\label{13}
S^{Sch}_{j\sigma}\equiv -2e(I^+_{j\sigma}+I^-_{j\sigma})=
2e^2\left[\Gamma^+_{j\sigma}p_0+\Gamma^-_{j\sigma}p_{\sigma}\right]
\end{equation}
is the high frequency ($\omega\to \infty$) limit of the shot-noise
(the Schottky noise), which is the sum of the components
corresponding to the tunneling current flowing to and from the
particle. The frequency dependent part is expressed as~\cite{kor}
%\begin{widetext}
\begin{eqnarray}\label{14}
S^c_{I_{j\sigma}I_{j'\sigma'}}(\omega)=2e^2(-1)^{j-j'}\{[
\Gamma^+_{j\sigma}G_{0\sigma'}(\omega)-
\Gamma^-_{j\sigma}G_{\sigma\sigma'}(\omega)]\Gamma^+_{j'\sigma'}p_0
+[\Gamma^+_{j'\sigma'}G_{0\sigma}(-\omega)-
\Gamma^-_{j'\sigma'}G_{\sigma'\sigma}(-\omega)]\Gamma^+_{j\sigma}p_0\nonumber\\
+[\Gamma^-_{j\sigma}G_{\sigma0}(\omega)-\Gamma^+_{j\sigma}G_{00}(\omega)]
\Gamma^-_{j'\sigma'}p_{\sigma'}
+[\Gamma^-_{j'\sigma'}G_{\sigma'0}(-\omega)-\Gamma^+_{j'\sigma'}
G_{00}(-\omega)]\Gamma^-_{j\sigma}p_{\sigma}\}\;.
\end{eqnarray}
%\end{widetext}
The shot noise of the total current (including the displacement
currents as well) is given by
\begin{eqnarray}\label{8c}
S_{II}=\sum_{j,j'}\frac{C^2_1C^2_2}{C^2C_jC_{j'}}\sum_{\sigma,\sigma'}
[\delta_{jj'}\delta_{\sigma\sigma'}S^{Sch}_{j\sigma}+S^c_{I_{j\sigma}I_{j'\sigma'}}(\omega)]\;.
\end{eqnarray}

\section{Results}

The analysis of the results we begin from a simplified situation,
when the electrodes are made of paramagnetic metals. Next the
device with ferromagnetic electrodes is considered. Since Coulomb
interactions break the electron-hole symmetry, one can expect that
the characteristics of the device for $E_0<E_F$ are different from
those for $E_0>E_F$. Therefore, the both situations are considered
separately.

\subsection{Paramagnetic case}

 In the system with paramagnetic electrodes both the
 channels for electrons with the spin $\uparrow$ and $\downarrow$
 are equivalent and the tunneling rates
$\gamma_{j\uparrow}=\gamma_{j\downarrow}=\gamma_{j}$,
$\gamma_{\uparrow}=\gamma_{\downarrow}=\gamma$,
$\Gamma^{\pm}_{\uparrow}=\Gamma^{\pm}_{\downarrow}=\Gamma^{\pm}$.
The total current $I_{1} =
-2e(f^+_1-f^+_2)\gamma_{1}\gamma_{2}/[\gamma+\Gamma^+]$ differs
form that for noninteracting electrons by the factor $\Gamma^+$ in
the denominator, which results form Coulomb interactions. In low
temperatures there is a current blockade for the voltage within
the range $-C/C_2<eV/(E_0-E_F)<C/C_1$. Dynamics of the
fluctuations are characterized by the eigenvalues
$\lambda_+=-\gamma+\Gamma^+$ and $\lambda_-=-\gamma-\Gamma^+$.
Using Eqs.(\ref{12})-(\ref{14}) and (\ref{8})-(\ref{9}) one can
derive the correlation function between the currents for electrons
with the same spin as
\begin{eqnarray}\label{19}
S_{I_{1\uparrow}I_{1\uparrow}}(\omega)=
2e^2\frac{\gamma_{1}(f^+_1\Gamma^-+f^-_1\Gamma^+)}
{\gamma+\Gamma^+}\nonumber\\
-2e^2\frac{\gamma_{1}^2}{\gamma+\Gamma^+}\left[\frac{f^+_1f^-_1
(\Gamma^-)^2}{\omega^2+\lambda_+^2}-\frac{
a_-}{\omega^2+\lambda_-^2}\right]
\end{eqnarray}
and between the different spins
\begin{equation}\label{20}
S_{I_{1\uparrow}I_{1\downarrow}}(\omega)=
2e^2\frac{\gamma_{1}^2}{\gamma+\Gamma^+}\left[\frac{f^+_1f^-_1
(\Gamma^-)^2}{\omega^2+\lambda_+^2}+\frac{
a_-}{\omega^2+\lambda_-^2}\right]\;,
\end{equation}
where $a_-=(1+f^+_1)[f^+_1(\Gamma^{+2}+2\gamma
\Gamma^+-\gamma^2)-2(\Gamma^+)^2]$. Thus, the power spectrum of
the total current through the left junction is expressed by
\begin{eqnarray}\label{21}
S_{I_{1}I_{1}}(\omega)=
4e^2\frac{\gamma_{1}(f^+_1\Gamma^-+f^-_1\Gamma^+)}
{\gamma+\Gamma^+}\nonumber\\
+8e^2\frac{\gamma_{1}^2}{\gamma+\Gamma^+}\frac{
a_-}{\omega^2+\lambda_-^2}\;.
\end{eqnarray}
The noise corresponding to the eigenvalue $\lambda_+$ is
completely cancelled. In the high-voltage regime the above
formulae are much simpler, e.g. for $V>0$ the current~\cite{gla}
$I_1=2e\gamma_{1}\gamma_{2}/(\gamma_1+2\gamma_2)$ and the Fano
factor~\cite{naz} \begin{equation}\label{21a}
 {\cal F}_{11}\equiv
\frac{S_{I_1I_1}(\omega=0)}{2eI_1} =
1-\frac{4\gamma_1\gamma_2}{(\gamma_1+2\gamma_2)^2}\;.
\end{equation}
(see also [\onlinecite{bb}] and references therein).

 For comparison in the case of noninteracting electrons ($E_c=0$)
 there are two independent
channels and the power spectrum can be written as
\begin{eqnarray}\label{16}
S^0_{I^0_{1\sigma}I^0_{1\sigma}}(\omega)=
2e^2\frac{\gamma_{1\sigma}[f^+_1\Gamma^-_{\sigma}+f^-_1\Gamma^+_{\sigma}]
}{\gamma_{\sigma}}\nonumber\\
-4e^2\frac{\gamma_{1\sigma}^2[f^+_1
(\Gamma^-_{\sigma})^2+f^-_1(\Gamma^+_{\sigma})^2]}{\gamma_{\sigma}
(\omega^2+\gamma_{\sigma}^2)}
\end{eqnarray}
for electrons with spin $\sigma$. For the paramagnetic electrodes
and in the limit of a large positive $V$ one gets [from
Eqs.(\ref{5b}) and (\ref{16})] the total current
$I^0_1=2e\gamma_{1}\gamma_{2}/(\gamma_1+\gamma_2)$ and the Fano
factor ${\cal
F}^0_{11}=1-2\gamma_{1}\gamma_{2}/(\gamma_{1}+\gamma_{2})^2$.~\cite{bb}

Let us present also the correlation function between the currents
through different tunnel junctions
\begin{eqnarray}\label{21c}
\mbox{Re}[S_{I_{1}I_{2}}(\omega)]=
4e^2\frac{\gamma_1\gamma_2}{\gamma+\Gamma^+}\frac{b_{12}}
{\omega^2+\lambda_-^2}\;,
\end{eqnarray}
where $b_{12}=\Gamma^{+2}(4+f^+_1+f^+_2-2f^+_1f^+_2)
+(f^+_1+f^+_2+2f^+_1f^+_2)(\Gamma^--\Gamma^+)\gamma$.
 Now, using Eq.(\ref{8c}) one gets the total power spectrum of the
device
%\begin{widetext}
\begin{eqnarray}\label{21d}
S_{II}(\omega)=
4e^2\frac{(C_2^2\gamma_1f^+_1+C_1^2\gamma_2f^+_2)\Gamma^-
+(C_2^2\gamma_1f^-_1 +C_1^2\gamma_2f^-_2)\Gamma^+}
{C^2(\gamma+\Gamma^+)}\nonumber\\
+8e^2\frac{\Gamma_{12}^2(\Gamma^{+2}+2\gamma\Gamma^+-\gamma^2)
-\Gamma_{12}\gamma_{12}\Gamma^{-2}-2\gamma_{12}^2\Gamma^{+2}}{(\gamma+\Gamma^+)
(\omega^2+\lambda_-^2)}\;,
\end{eqnarray}
%\end{widetext}
where $\Gamma_{12}=(C_2\gamma_1f^+_1-C_1\gamma_2f^+_2)/C$,
$\gamma_{12}=(C_2\gamma_1-C_1\gamma_2)/C$. One can check that in
the zero-frequency limit
$S_{I_{1}I_{1}}(0)=$Re$[S_{I_{1}I_{2}}(0)]=S_{I_{2}I_{2}}(0)$.
Therefore, the Fano factors ${\cal F}_{11}={\cal F}_{12}={\cal
F}_{22}$, which in the high-voltage range can be simply expressed
as ${\cal F}=1-4\gamma_1\gamma_2/(\gamma_1+2\gamma_2)^2$.

 We are also
interested in charge and spin fluctuation induced by the flowing
current. Using the formula (\ref{10}) and (\ref{11}) for the
paramagnetic device one can write the charge-charge and the
spin-spin correlation function as
\begin{eqnarray}\label{22}
S_{NN}(\omega)=
\frac{8e^2\Gamma^+\Gamma^-}{(\gamma+\Gamma^+)(\omega^2+\lambda_-^2)}\\
S_{MM}(\omega)=
\frac{2\mu_B^2\Gamma^+\Gamma^-}{(\gamma+\Gamma^+)(\omega^2+\lambda_+^2)}\;.
\end{eqnarray}
From a frequency dependence of the correlation functions $S_{NN}$
and $S_{MM}$ one can assign the relaxation time corresponding to
the charge and the spin fluctuations as
$\tau_{charge}=-1/\lambda_-$ and $\tau_{spin}=-1/\lambda_+$,
respectively. One can check that the same result for the
correlation functions can be derived from the two-level
generation-recombination approach~\cite{vliet} using
$S_{XX}(\omega)=4\mbox{var}(X)\tau/(\omega^2\tau^2+1)$, where
var$(X)=\langle X^2\rangle-\langle X\rangle^2$ is the variance of
the quantity $X$. Since $\tau_{spin} > \tau_{charge}$ then spin
fluctuations occur in a low frequency regime, while the charge
fluctuations in higher frequencies. The amplitude of the spin
noise $S_{MM}(\omega=0)$ is larger than $S_{NN}(\omega=0)$ (in
some cases the difference can be a few orders of
magnitudes~\cite{prb}). In the paramagnetic system the spin
fluctuations, however, do not contribute to the current shot
noise. The frequency dependence of the power spectrum (\ref{21})
has then a Lorentzian form with the relaxation time
$\tau_{charge}$.

\subsection{Ferromagnetic electrodes and $E_0 < E_F$}

Let us first consider the ferromagnetic double tunnel barrier
device, in which the particle level is below the Fermi level of
the electrodes. A typical voltage dependence of the current is
shown in Fig.1a. The $I$-$V$ function has a step like shape, with
the current blockade for small voltages [in the range
$-C/C_1<eV/(E_F-E_0)<C/C_2$] and the plateaux in the limit of
large voltages, in which
\begin{eqnarray}\label{14a}
I_1=\left\{\begin{array}{lcc}
e\frac{\gamma_{1\uparrow}\gamma_{1\downarrow}(\gamma_{2\uparrow}+
\gamma_{2\downarrow})}
{\gamma_{\uparrow}\gamma_{\downarrow}-\gamma_{2\uparrow}\gamma_{2\downarrow}}
&\mbox{for}&V\gg (E_F-E_0)/e,\;\;\\
-e\frac{\gamma_{2\uparrow}\gamma_{2\downarrow}(\gamma_{1\uparrow}+
\gamma_{1\downarrow})}
{\gamma_{\uparrow}\gamma_{\downarrow}-\gamma_{1\uparrow}\gamma_{1\downarrow}}
&\mbox{for}&V\ll -(E_F-E_0)/e.\;\;\end{array}\right.
\end{eqnarray}
We remind that according to our assumptions $|V|\ll \Delta E, E_c$
and the tunneling rates $\gamma_{j\sigma}$ are independent of $V$,
even for the so called {\it high voltages} when the currents are
given by Eq.(\ref{14a}). The current intensities (\ref{14a}) for
large positive and negative voltages are different, in contrast to
the case of noninteracting electrons, where both the $I$-$V$ steps
are equal. Fig.1a shows that the height of the steps depends on
the magnetic asymmetry of the electrodes, and an increase of the
magnetic polarization $P_1$ in the left electrode reduces the
current for $V>0$. If this electrode is made of a half-metallic
ferromagnet (i.e. for $P_1=1$ and $\gamma_{1\downarrow}=0$) the
conducting channel corresponds only to electrons with the spin
$\uparrow$, and there is the Coulomb blockade $I_1=0$ in low
temperatures ($k_BT\ll (E_F-E_0)$) for any positive voltage. An
electron with the spin $\downarrow$, which has tunneled form the
right electrode into the particle, is captured there forever. The
electron can neither tunnel to the left nor to the right
electrode, and blocks the conducting channel for electrons with
the spin $\uparrow$. Electronic transport can only occur for large
negative voltages. Such the device works as a diode.

Using Eq.(\ref{14a}) one finds
\begin{equation}\label{14a1}
TMR=\frac{4\alpha P_1P_2}{1-P_1^2+2\alpha-2\alpha P_1P_2}
\end{equation}
in the limit $V\gg(E_F-E_0)/e$. For comparison the value for
noninteracting electrons
\begin{equation}\label{14a2}
TMR^0=\frac{4\alpha P_1P_2}{1-P_1^2+2\alpha-2\alpha
P_1P_2+\alpha^2(1-P_2^2)}\;,
\end{equation}
is much smaller, especially in the system with asymmetric tunnel
junctions ($\alpha\gg 1$). One can say that Coulomb interactions
enhance the value of TMR.

 The power spectrum on the
conducting step (for $V>0$) is given by
\begin{eqnarray}\label{14b}
S_{I_{1}I_{1}}(\omega)=
2e^2\frac{\gamma_{1\uparrow}\gamma_{1\downarrow}(\gamma_{2\uparrow}+\gamma_{2\downarrow})}
{\gamma_{\uparrow}\gamma_{\downarrow}-\gamma_{2\uparrow}\gamma_{2\downarrow}}
\nonumber\\
-\frac{4e^2\gamma_{1\uparrow}\gamma_{1\downarrow}(\gamma_{2\uparrow}
+\gamma_{2\downarrow})}{\Delta
(\gamma_{\uparrow}\gamma_{\downarrow}-\gamma_{2\uparrow}\gamma_{2\downarrow})^2}
\sum_r  r\lambda_r\frac{\lambda_r a + b}{\omega^2+\lambda^2_r}
\end{eqnarray}
where
$a=-\gamma_{1\uparrow}\gamma_{1\downarrow}(\gamma_{2\uparrow}+\gamma_{2\downarrow})$
and
$b=\gamma_{1\uparrow}^2\gamma_{2\downarrow}(\gamma_{2\uparrow}-\gamma_{1\downarrow})+
\gamma_{1\downarrow}^2\gamma_{2\uparrow}(\gamma_{2\downarrow}-\gamma_{1\uparrow})-2\gamma_{1\uparrow}\gamma_{1\downarrow}\gamma_{2\uparrow}\gamma_{2\downarrow}$.
 The eigenvalue in this case is $\lambda_r=(-\gamma_{\uparrow}-\gamma_{\downarrow}+r\Delta)/2$,
 and $\Delta=\sqrt{(\gamma_{\uparrow}-\gamma_{\downarrow})^2+4\gamma_{2\uparrow}\gamma_{2\downarrow}}$.
 The voltage dependence of the Fano factor is presented in Fig.1b. One
 can show  that the zero-frequency power spectrum
 $S_{I_{j}I_{j'}}(\omega=0)$  corresponding
 to the currents through different tunnel
 junctions are equal, and thus, the Fano factors
 ${\cal F}_{11}={\cal F}_{12}={\cal F}_{22}$ for any
 model parameters (for any transition rates $\gamma_{j\sigma}$ at
 any voltage).  In the regime of high-voltage its value is
 ${\cal F}=1+2b/(\gamma_{\uparrow} \gamma_{\downarrow}-\gamma_{2\uparrow}\gamma_{2\downarrow})^2$.
 If the coefficient $b$ is negative, then ${\cal F}<1$ and the noise
 is of the sub-Poissonian type.
 It occurs for $2\alpha P_1^2(1-P_2^2)<(1-P_1P_2)(1-P_1^2)$. The
 transition from the sub-Poissonian to the super-Poissonian type
 of the current shot noise is a continuous process. In order to
 understand it we plotted in Fig.2 the frequency dependent part of the
 correlation functions $S^c_{I_{1\sigma}I_{1\sigma'}}(\omega=0)$
 [given by Eq.(\ref{14})] for the currents of electrons with different
 spins through the left junction in the high-voltage limit.
One can expect competition between tunneling processes for
electrons with the spin
 $\uparrow$ and $\downarrow$, which leads to an enhancement
 of the current noise. For simplicity the right electrode
 is taken paramagnetic, i.e. the source electrode can emit
 electrons with the same
 transition rate ($\gamma_{2\uparrow}=\gamma_{2\downarrow}$).
 The drain electrode is ferromagnetic and therefore, there is
 an asymmetry between the out-going channels for electrons with
 opposite spin directions, which is described by the
 magnetic polarization $P_1$. For $P_1=0$, the functions are equal
 $S^c_{I_{1\uparrow}I_{1\uparrow}}(0)=S^c_{I_{1\downarrow}I_{1\downarrow}}(0)
 =S^c_{I_{1\uparrow}I_{1\downarrow}}(0)$ and negative. It means
 that all tunneling events are anti-correlated, which leads to a
 reduction of the noise. An increase of the polarization $P_1$
 increases the tunneling rate $\gamma_{1\uparrow}$ for electrons
 with the spin $\uparrow$, they can faster leave the
 particle. Electrons with the opposite spin ($\downarrow$)
 spend a long time on the particle. It effects in the spin
 accumulation,~\cite{fsette} which is responsible for an increase
 of $S^c_{I_{1\uparrow}I_{1\uparrow}}(0)$ and
$S^c_{I_{1\uparrow}I_{1\downarrow}}(0)$. Their values can cross
zero and achieve maxima for $P_1\to 1$. The function
$S^c_{I_{1\downarrow}I_{1\downarrow}}(0)$ is always negative (for
$P_1>0$). The process results an enhancement of the shot noise and
the transition to the super-Poissonian range. The maximum value of
the Fano factor ${\cal F}=
1+2\gamma_{2\uparrow}/\gamma_{1\downarrow}$ occurs for the left
electrode made of a half-metallic ferromagnet ($P_1=1$). Fig.2
shows also that a large asymmetry factor $\alpha\gg 1$ between the
left and the right tunnel barrier can prefer the transition to the
super-Poissonian shot noise (see the dashed curves corresponding
to $\alpha=10$).

\subsection{Ferromagnetic electrodes and $E_0 > E_F$}

In the case of $E_0>E_F$ one can expect similar characteristics of
our device to those presented above for $E_0<E_F$. It is really
the case, but only for the high-voltage regime, where the $I$-$V$
curve has plateaux, whose level is given by Eq.(\ref{14a}). Fig.3
presents the voltage dependence of the current and the Fano
factor. (Since the curves in the range of negative $V$ are very
similar to those from Fig.1, we present the dependences for $V>0$
only). It is seen a resonant-like peak of the current in the range
of moderate voltages, at $E_0-eV_2\approx E_F$ (i.e. for
$V/(|E_0-E_F|/e)\approx 2$ in Fig.3a). Its height can be much
above the plateau level in the device with large asymmetry of the
tunnel junctions. The most pronounced peak is for the device with
the left electrode made of a half-metallic ferromagnet ($P_1=1$,
$\gamma_{1\downarrow}=0$). The total current, in this case, can be
written as
\begin{equation}\label{15b}
I_{1\uparrow}=e\frac{(f^+_1-f^+_2)f^-_2\gamma_{1\uparrow}\gamma_{2\uparrow}}
{(1-f^+_1f^+_2)\gamma_{1\uparrow}+(1-f^{+2}_2)\gamma_{2\uparrow}}\;.
\end{equation}
It is worth noticing, that in this limit the current (\ref{15b})
and the occupation probability $p_{\uparrow}$, $p_{\downarrow}$,
$p_0$ are independent of the transition rate
$\gamma_{2\downarrow}$. The current peak is the resonant-like
transition of electrons through the particle level and the current
blockade effect in the low and the high-voltage range. For a small
$V$ the position of the particle level $E_0-eV_2$ is above the
Fermi level $E_F$ of the right electrode and electrons can not
tunnel to the particle, whereas in a high-voltage range there is a
Coulomb blockade of the conducting channel by an electron with
spin $\downarrow$ captured on the particle. The width of the
current peak depends on the smearing of the Fermi surface and
decreases with a decreasing temperature.

 The $I$-$V$
curve (\ref{15b}) resembles that obtained in the case of resonant
tunneling through double barrier in semiconductors (the Esaki
diode).~\cite{esaki} The nature of the both tunneling effects is,
however, different. In the present case the negative differential
resistance (NDR) is caused by Coulomb interactions between
electrons on the particle (by the Coulomb blockade effect). In the
Esaki diode~\cite{esaki} the charge accumulation in the well is
irrelevant for electronic transport and the NDR results from a
shift of the conduction band of the source electrode out of the
resonant tunneling range (see
[\onlinecite{azbel,flub,flui,fluk,flubb,bb}], which considered
coulomb interactions in resonant tunneling as well). The width of
the peak depends in the Esaki diode on the electronic structure of
the device. It can be smeared due to fluctuations of the bottom of
the potential well.~\cite{flubb,bb} In our model the position of
$E_0$ is fixed and the broadening of the peak results only from
the thermal distribution of electrons around the Fermi level.

Fig.3b shows the voltage dependence of the Fano factor. Its value
is below unity in the low-voltage range and rapidly increases when
$E_0-eV_2$ crosses the Fermi level $E_F$ (i.e. for
$V/(|E_0-E_F|/e)>2$ in Fig.3b). The increase of ${\cal F}$ is only
in a narrow range of $V$, in the same in which the NDR effect is
observed. In the high-voltage regime the noise is super-Poissonian
for the most situations exhibited in Fig.3b. The voltage
dependence of the Fano factor in the present case (see the curve
for $P_1=1$ in Fig.3b) is qualitatively different from that in the
resonant tunneling diode~\cite{flub,flui,fluk}, where ${\cal F}$
shows a large peak in the NDR region. The origin of the Fano peak
is activation of interaction-induced fluctuations of the band
bottom in the quantum well, when the system passes to the
off-resonant electronic transport.~\cite{flui,fluk,flubb,bb} As we
have explained already in the previous section, the high value of
${\cal F}$ in our system is related with the asymmetry of the
conducting channels for electrons with the opposite spin
directions.

Flowing electrons induce the charge and the spin fluctuations on
the particle with the characteristic frequencies
$1/\tau_{charge}=-\lambda_-$ and $1/\tau_{spin}=-\lambda_+$,
respectively. These fluctuations should be seen in the current
noise. Therefore, we separate the Schottky term $S^{Sch}_I$ from
the current noise and perform the spectral decomposition of the
frequency dependent part $S^c_{II}(\omega)$. The total power
spectrum can be expressed as
\begin{equation}
S_{II}(\omega)=S^{Sch}_I+S^{c+}_{II}(\omega)+S^{c-}_{II}(\omega)\;,
\end{equation}
where $ S^{Sch}_I=(C_2^2S^{Sch}_{I1}+C_1^2S^{Sch}_{I2})/C^2 $ and
\begin{equation}
S^{c\pm}_{II}(\omega)=\pm\left(\frac{C_1C_2}{C}\right)^2
\sum_{j,j'}\frac{\lambda_{\pm}}{C_jC_{j'}}\frac{\lambda_{\pm}a_{jj'}+
b_{jj'}}{\omega^2+\lambda_{\pm}^2}\;.
\end{equation}
The coefficient $a_{jj'}$ and $b_{jj'}$ are determined from
Eq.(\ref{14}), (\ref{8}), (\ref{9}) and (\ref{3}). The voltage
dependences of these terms for $\omega=0$ are presented in Fig.4.
The system is the same as studied above (Fig.3), in which the
asymmetry between the tunnel barriers is $\alpha=10$. The
transition rates $\gamma_{2\sigma}$ are larger than
$\gamma_{1\sigma}$, and therefore  $S^{Sch}_{I2}>S^{Sch}_{I1}$. In
the low-voltage range the term $S^{Sch}_{I2}$ is very large and
dominates in $S^{Sch}_{I}$. The Fano factor ${\cal
F}=[S^{Sch}_I+S^{c+}_{II}(0)+S^{c-}_{II}(0)]/(2eI)$ is, however,
below unity. In the considered system we change the magnetic
polarization $P_1$, which influences of $S^{Sch}_{I1}$, but it is
irrelevant for $S^{Sch}_{I}$. It explains, why all the curves in
Fig.4a are so close to each other.

Fig.4b and 4c show the terms $S^{c+}_{II}(0)$ and $S^{c-}_{II}(0)$
corresponding to the contribution of the spin and the charge
fluctuations to the current noise. They are negative in the
low-voltage range and positive for larger voltages. This indicates
a change of current correlations when the particle level crosses
the Fermi level  ($E_0-eV_2\approx E_F$). The value
$S^{c+}_{II}(0)$ strongly increases with an increase of the
magnetic polarization $P_1$. Since $S^{c-}_{II}$ and $S^{Sch}_I$
(see Fig.4c and 4a) are weakly dependent on $P_1$, it is evident
that $S^{c+}_{II}$ is responsible for an enhancement of the Fano
factor. Frequency dependent measurements of the current noise can
confirm our prediction, that low frequency fluctuations dominate
in the super-Poissonian noise in ferromagnetic tunnel junctions.

\section{SUMMARY}

 Summarizing, our sequential tunneling studies,
performed in the ferromagnetic double barrier device with the
atomic size particle, showed a few interesting effects. First,
Coulomb interactions lead to an enhancement of the TMR effect.
Second, an electron-hole symmetry is broken in the system, due to
Coulomb interactions. The characteristics of the ferromagnetic
device, with the electronic state $E_0$ of the spacer particle
below the Fermi level $E_F$ of the electrodes, are qualitatively
different from those for the case of $E_0>E_F$. We showed that the
system, in which $E_0<E_F$ and one electrode is made of a
half-metallic ferromagnet, can operate as a diode. When $E_0>E_F$
the device showed the NDR effect, which is better pronounced for
ferromagnetic electrodes with different magnetic polarizations.
Third, the transition from the sub-Poissonian to the
super-Poissonian current noise is a continuous process, which
depends on the magnetic asymmetry between the tunneling channels
for electrons with the spin $\uparrow$ and $\downarrow$. The
asymmetry between the left and the right tunnel barrier can
facilitate the transition to the super-Poissonian range. Spin
fluctuations are relevant for the super-Poissonian current noise
and they are activated in the Coulomb blockade regime. The charge
fluctuations are responsible for the sub-Poissonian current noise.
The spin and the charge fluctuations have distinct relaxation
times $\tau_{spin}>\tau_{charge}$, which can be observed in
frequency dependent measurements of the power spectrum in a low
and in a high-frequency range, respectively.

\acknowledgments{
    The paper is supported from the State Committee for Scientific Research
Republic of Poland within Grant No.~2 P03B 075 14.}

\newpage
\begin{figure}
\epsfxsize=12cm \epsffile{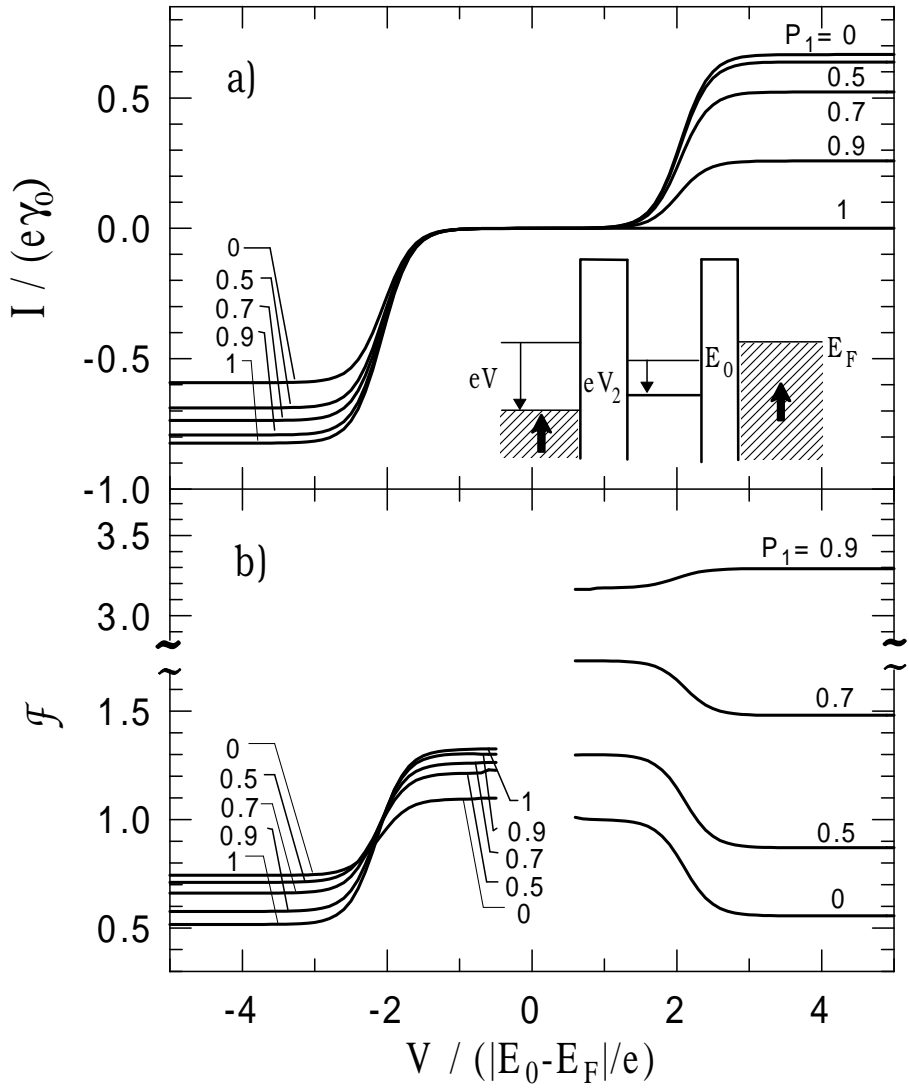 }\vskip 1cm  \caption{Voltage
dependence of the current (a) and the Fano factor (b) in the
ferromagnetic double barrier with the resonating level $E_0<E_F$
for different magnetic polarizations of the left electrode $P_1 =
0$, 0.5, 0.7, 0.9, 1. The polarization of the right electrode is
$P_2=0.4$, the asymmetry between the barrier $\alpha=1$, the
capacitances $C_1=C_2$, the difference $|E_F-E_0|/e$ is taken as
unity, and the temperature $T=0.1$. The inset shows the scheme of
the electronic structure.}
\end{figure}

\newpage
\begin{figure}
\vskip 3cm \epsfxsize=12cm \epsffile{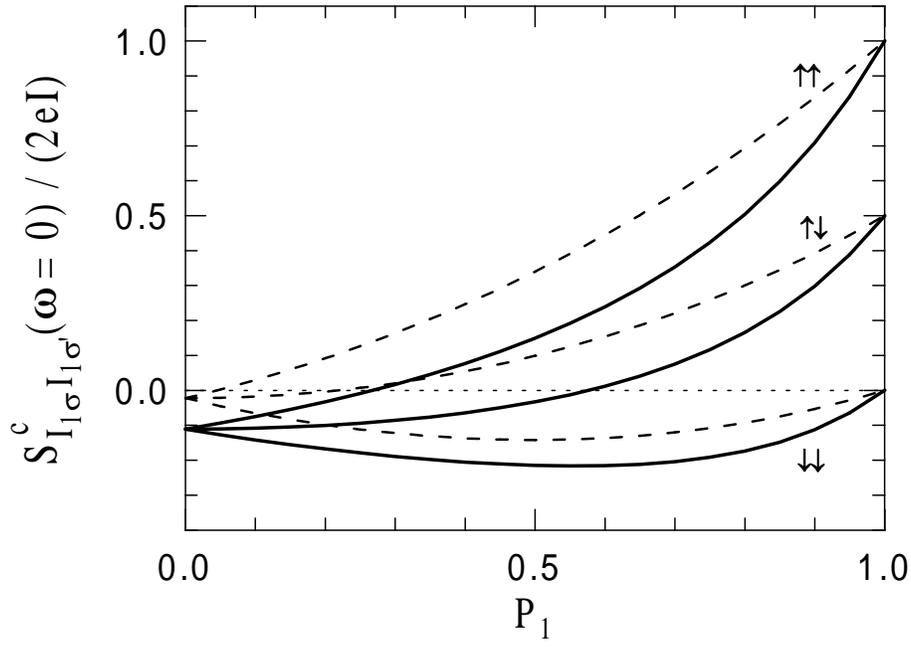} \vskip 1cm
\caption{The frequency dependent part
 of the correlation functions $S^c_{I_{1\uparrow}I_{1\uparrow}}$,
$S^c_{I_{1\downarrow}I_{1\downarrow}}$ and
$S^c_{I_{1\uparrow}I_{1\downarrow}}$ at $\omega=0$
 for the currents with different
spin orientation  as a function of the magnetic polarization $P_1$
in the left electrode. The plot was done for the currents in the
high-voltage limit and for the device with the right paramagnetic
electrode ($P_2=0$), the asymmetry between the tunnel barriers is
taken as $\alpha=1$ (solid curves) and $\alpha=10$ (dashed
curves).}
\end{figure}

\newpage
\vskip 1cm\epsfxsize=12cm \epsffile{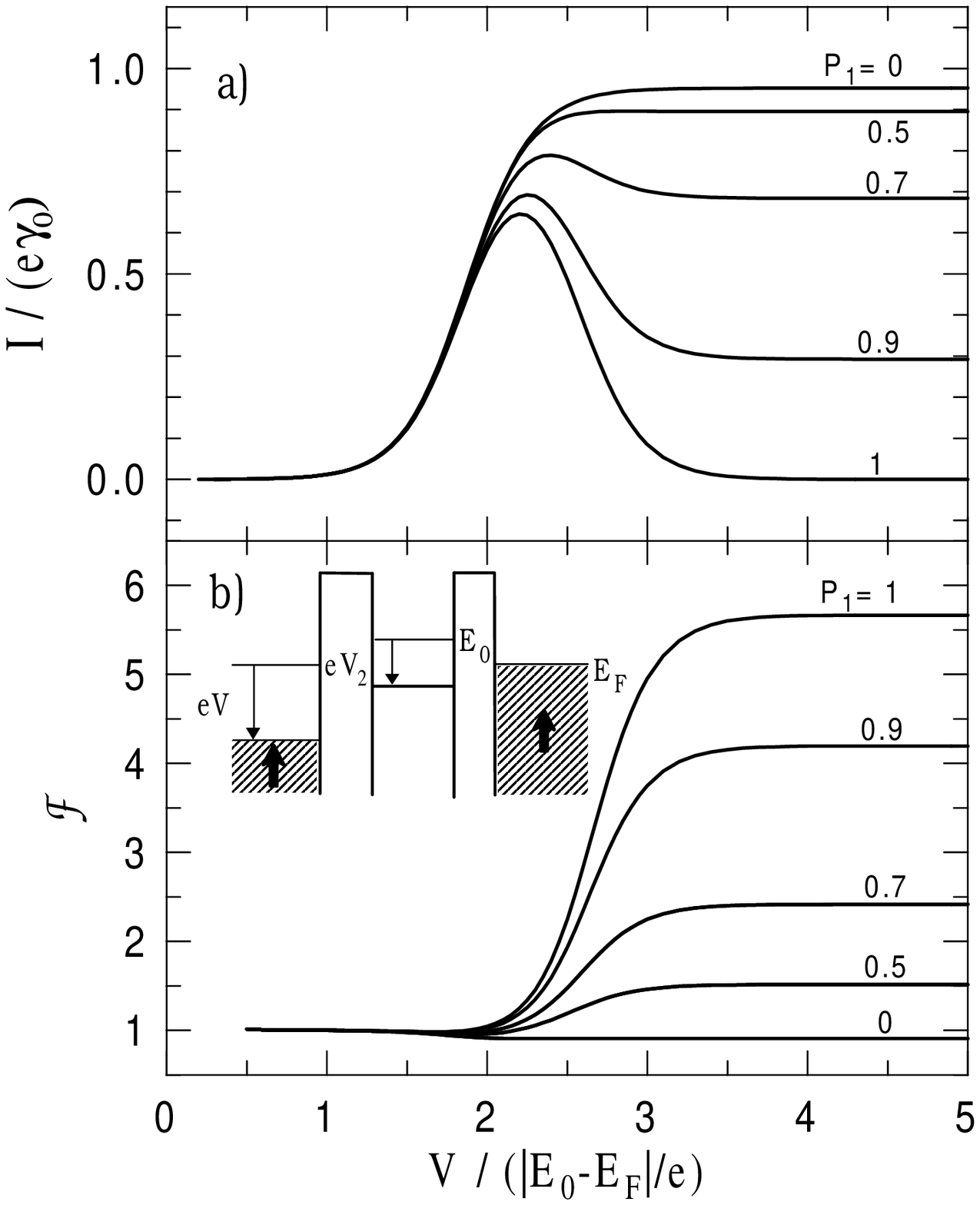 }\vskip 1cm
\begin{figure}
\caption{Voltage dependence of the current (a) and the Fano factor
(b) in the ferromagnetic double barrier with the resonating level
$E_0>E_F$ for different magnetic polarizations of the left
electrode $P_1 = 0$, 0.5, 0.7, 0.9, 1. The polarization of the
right electrode is $P_2=0.4$, the asymmetry between the barrier
$\alpha=10$, the capacitances $C_1=C_2$, the difference
$|E_F-E_0|/e$ is taken as unity, and the temperature $T=0.1$. The
inset shows the scheme of the electronic structure for this case.}
\end{figure}

\newpage
\epsfxsize=12cm \epsffile{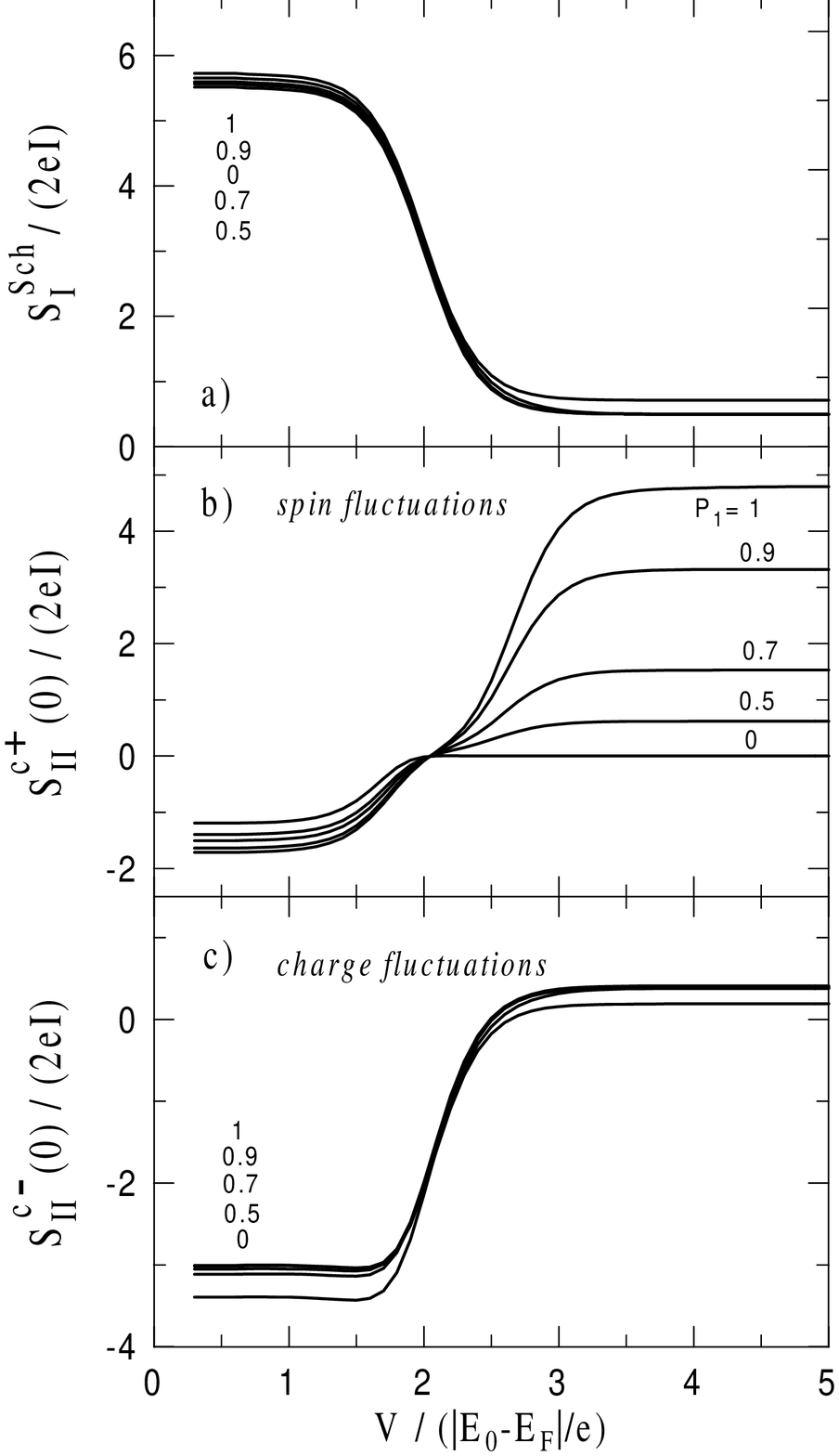}
\begin{figure}
\caption{Voltage dependence of the different components of the
zero-frequency current noise: the Schottky term (a), the frequency
dependent parts $S^{c+}_{II}$ and $S^{c-}_{II}$ corresponding to
the relaxation time $\tau_{spin}=-1/\lambda_+$ (b) and
$\tau_{charge}=-1/\lambda_-$ (c), respectively. The plots are done
for the ferromagnetic device the same as in Fig.3 with the
magnetic polarization of the left electrode $P_1 = 0$, 0.5, 0.7,
0.9, 1.}
\end{figure}

\end{document}